\documentstyle [prl,aps] {revtex}
\begin{document}
\draft
\title{ PHOTOFISSION AND QUASI-DEUTERON-NUCLEAR STATE AS MIXING 
OF BOSONS AND FERMIONS
\thanks{To appear in Zeit. f. Phys. A}
}
\author{ G. Kaniadakis, A. Lavagno
\thanks{e-mail: alavagno@polito.it} 
and P. Quarati}
\address{Dipartimento di Fisica - INFM - Politecnico di Torino - 
10129 Torino, Italy \\ 
Istituto Nazionale di Fisica Nucleare - Sezione di Cagliari and 
Sezione di Torino}

\maketitle
\begin {abstract} 
The empirical-phenomenological quasi-deuteron photofission description is 
theoretically 
justified within the semiclassical, intermediate statistics model. 
The transmutational 
fermion (nucleon) - boson (quasi-deuteron) potential plays an 
essential role 
in the present context and  
is expressed in terms of thermodynamical and of microscopical quantities,  
analogous to those commonly used in the superfluid nuclear model.
\end {abstract}

\section{Introduction}

Over the past several years, many efforts have been directed 
to develop methods for the physical 
description of complex many-body systems. 
The implementation of quantum approaches like the Hartree-Fock, 
the Random Phase Approximation  and the Time Dependent Hatree-Fock 
Method have allowed the explanation of a great variety of experimental 
results and particularly of the properties of the nuclear ground 
state  [1,2].\\
In the literature, 
the nuclear dynamics is described by different theories. 
In the most cases, approaches based on semiclassical one-body equations 
(Fokker-Plank, 
Boltzmann, Boltzmann-Langevin equations) are use. These methods can be 
regarded as semiclassical analogs 
of the mentioned quantum approaches [3,4] and include the Pauli exclusion 
and the inclusion principle [5] 
to represent correctly systems of fermions, bosons or anyons.

In several problems in condensed matter and in nuclear physics, 
the correlation effects between pairs of 
fermions are quite relevant in the interpretation of  
experimental results. 
Similarly, the interactions among bosons are relevant 
to the superfluid nuclear model, 
the interacting boson model and the mean field boson 
approximation and allow the explanation of many collective nuclear properties. 
In these cases, the interaction 
among the valence nucleons outside the core produces pairs of correlated 
nucleons that can be approximated with bosonic particles (like, for instance, 
the quasi-deuterons). 
The pairing of fermions leads a certain amount of nucleons to a transition 
from one statistics to another (e.g. Bose-Einstein). 
This transition can be modeled with 
a transmutational potential (TP) and implies that a nucleon can 
be considered, in particular excited states, 
as an ensemble of fermions and bosons or as an ensemble of particles 
of intermediate statistics.

Semiclassical kinetic approaches to intermediate statistics [5-7] are  
based on the classical Fokker-Planck equation and describe the time 
evolution of the distribution function $n(t,{\bf v})$. 
The Fokker-Planck equation is corrected with the 
insertion of the factor $1+\kappa n(t,{\bf v})$ into the transition rates,  
where $\kappa$ varies between $-1$ 
(Fermi-Dirac distribution (FD)) and $+1$ (Bose-Einstein distribution (BE)), 
($\kappa=0$ is the case of the Maxwell-Boltzmann distribution). 
The effect on the distribution function of the exclusion ($\kappa<0$) 
and of the inclusion ($\kappa>0$) principle, 
can be evaluated with this approach.\\
This approach is 
not unique, different non linear expressions can also be considered [7]. 
However, more involved expressions are not required for the 
discussion presented in this work.\\
In the present paper, 
we call statistical transmutation the variation of the particle state 
from its characteristic value $\kappa$ to a new value $\kappa{'}$ (this 
definition includes, in particular, the transmutation from $\kappa=-1$ to 
$\kappa{'}=+1$). 

To describe the importance of the TP,  
we fix our attention to the photofission of heavy nuclei in the energy 
range 
of excitation between $40$ $MeV$ and $100$ $MeV$ (usually called 
quasi-deuteron (QD) energy region [8,9]).\\
In photonuclear reactions, it is well known that, above the GDR and 
below the pion photoproduction threshold, the photoabsorption involves 
two nucleons to satisfy the kinematical conservation laws. 
This circumstance helps the formation of 
strongly correlated pairs of nucleons (quasi-deuteron states). The 
photoabsorption cross section is proportional to the deuteron 
photodisintegration cross section.\\
Recently, we have developed an extension of the 
QD model, to describe the total 
photoabsortion and the photofission of heavy nuclei [10-11]
(an extension to higher excitation energies, pion photoproduction region, 
of the QD photofission model 
has been recently developed by Arruda-Neto and collaborators [12,13]).
The agreement of the model with the experimental results is 
quite good for a wide variety of nuclei ($Bi$, $Th$, $U$) 
(see also Ref.s [14-16]). 
Our phenomenological-empirical model provides us useful  
quantities, such as the number of QD participating to photofission at different 
excitation energies. Its use, together with the semiclassical 
kinetic approach, provides a theoretical basis for the 
phenomenological model 
itself. The evaluation of the TP in analogy with the superfluid model, 
allows the introduction of an analytical expression for the excited QD nuclear 
state as a mixture of bosons and fermions. The TP can be given an 
interpretation in terms of the occupation operators. 
An interpretation of the TP in terms of thermodynamical quantities is also 
given.

In Sect. II, we describe the equations of the transmutational kinetic model 
and provide the free nucleon fraction and the probability of 
participation of one nucleon to a QD pair.\\
In Sect. III, we calculate the transmutational potential using  
the results of a phenomenological approach to photofission, valid 
in the QD energy region
and expressed in terms of the nuclear latent heat.\\
In Sect. IV, we give an interpretation of the transmutational 
potential using an analogy with  the superfluid model of the nucleus.
Conclusion are outlined in Sect. V.

\section{Nucleon-Quasi Deuteron transmutation model}

The protons and the neutrons are distinguished by their third 
isospin component $t_3=\tau$ ($\bar\tau=-\tau$). The nucleon mass 
is indicated by $m_{_N}=939$ $MeV$ and the occupational number in 
the velocity space by $n^\tau_{_N}(t,{\bf v})$. The QD particles, 
neglecting their binding energy, have mass $m_{_D}=2 \, m_{_N}$, 
$L=0$, $S=1$, $T=0$. Their occupational number is $n_{_D}(t,{\bf v})$.\\
By using an intermediate statistics with a linear enhancement or 
inhibition factor to calculate the QD photofission cross section, 
we can assign to the nucleons a value of the parameter $\kappa$,  
intermediate between $-1$ and $1$ and different from zero. 
When the excitation energy increases, the parameter $\kappa$ varies 
because the QD particles are more numerous as the nucleus is more 
excited. Therefore, the bosonic component of each nucleon increases  
and one can study the TP from the statistical transmutation from $\kappa$ to 
$\kappa^{'}$. We are interested in the transmutation nucleon-QD or viceversa, 
therefore we must fix, in this context, the states $\kappa$ and $\kappa^{'}$ 
as the fermion and boson state, respectively.\\
An alternative, but equivalent, procedure is 
to assume that the system (nucleus) is composed by a certain number of 
fermions (nucleons) 
and by a certain number of bosons (QD), depending on the excitation energy.
Therefore, we write two different kinetic 
equations for the nucleons, with $\kappa=-1$, and for the QD, with 
$\kappa=+1$. In this way it can be clearly understood the role of 
the fermion-boson TP.

We consider the energy as a continuous variable, the two 
functions $n^\tau_{_N}(t,{\bf v})$ and $n_{_D}(t,{\bf v})$ obey 
the following system of coupled equations [6]:

\begin{equation}
\frac {\partial n_{_N}^\tau ( t , {\bf v} )} {\partial t} + 
{\bf \nabla} {\bf j}_{_N}^\tau (t,{\bf v})+
j_{_{2N \leftrightarrow D}}^{\tau \bar\tau} (t,{\bf v})=0 \ \ ,
\end{equation}

\begin{equation}
\frac {\partial n_{_D} ( t , {\bf v} )} {\partial t} + 
{\bf \nabla} {\bf j}_{_D} (t,{\bf v})+
j_{_{D \leftrightarrow 2N}}^{\tau \bar\tau} (t,{\bf v})=0 \ \ .
\end{equation}
\noindent
These equations can be viewed as the continuity equations 
of a nucleon and of a quasi-deuteron gas in the velocity space. 
When the energy is a discrete variable, we can substitute 
Eq.s (1) and (2) with a system of coupled master equations.\\
In the velocity space, the currents ${\bf j}_{_N}^\tau (t,{\bf v})$ and 
${\bf j}_{_D} (t,{\bf v})$ have the following expressions:

\begin{equation}
{\bf j}_{_N}^\tau (t,{\bf v})=- \left [ {\bf J}_{_N}^\tau (t,{\bf v})+ 
{\bf\nabla}D_{_N}^\tau (t,{\bf v})\right ] 
n_{_N}^\tau (t,{\bf v}) [1-n_{_N}^\tau (t,{\bf v})]-
D_{_N}^\tau (t,{\bf v}) {\bf\nabla} n_{_N}^\tau(t,{\bf v}) \ \ ,
\end{equation}
\noindent

\begin{equation}
{\bf j}_{_D} (t,{\bf v})=- \left[ {\bf J}_{_D} (t,{\bf v})+
{\bf\nabla}D_{_D} (t,{\bf v})\right ] 
 n_{_D} (t,{\bf v}) [1+n_{_D} (t,{\bf v})]-D_{_N} (t,{\bf v}) {\bf\nabla} 
n_{_D} (t,{\bf v}) \ \ ,
\end{equation}
\noindent
where ${\bf J}_{_{N,D}} (t,{\bf v})$ and $D_{_{N,D}} (t,{\bf v})$ are, 
respectively, the drift and 
diffusion coefficients for the nucleon and the QD.\\ 
The net transmutational current 
$j_{_{2N \leftrightarrow D}}^{\tau \bar\tau}$ 
($j_{_{D \leftrightarrow 2N}}^{\tau \bar\tau}$)
takes into account 
the formation of a quasi-deuteron (two nucleons) from two nucleons 
(a deuteron) with different 
third isospin component; its expression, according 
to the exclusion-inclusion principle (EIP), is the following [6]:
\begin{eqnarray}
j_{_{2N \leftrightarrow D}}^{\tau \bar\tau}(t,{\bf v})=
r_{_{2N \rightarrow D}}(t) \, g_{_N}(t,{\bf v}) \, 
n_{_N}^\tau (t,{\bf v}) \, n_{_N}^{\bar\tau} (t,{\bf v}) \,
[1+n_{_D} (t,{\bf v})] \nonumber \\
-r_{_{D \rightarrow 2N}}(t) \, g_{_D}(t,{\bf v}) 
\, n_{_D}(t,{\bf v}) \, [1-n_{_N}^\tau (t,{\bf v})] 
[1-n_{_N}^{ \bar\tau} (t,{\bf v})] \ \ ,
\end{eqnarray}
\noindent 
where $r_{_{2N \rightarrow D}}(t)$ is the transmutation 
rate of two nucleons in a quasi-deuteron, $r_{_{D\rightarrow 2N}}(t)$ 
is the transmutation 
rate of a quasi-deuteron into two nucleons, $g_{_N}(t,{\bf v})$ 
and $g_{_D}(t,{\bf v})$ 
are functions taking into account the interaction among the 
identical particles belonging to the bound system we are considering. \\
From the definition of the net transmutational current we have:

\begin{equation}
j_{_{2N \leftrightarrow D}}^{\tau \bar\tau}(t,{\bf v})= 
-j_{_{D \leftrightarrow 2N}}^{\tau \bar\tau}(t,{\bf v}) 
\end{equation}
\noindent
Equations (1-6) define univocally 
the diffusion process of neutrons, protons and quasi-deuterons 
in the velocity space and the formation and disgregation of 
quasi-deuterons inside the nucleus.

The statistical distribution $n_{_{N,D}}({\bf v})$ can 
be obtained, in stationary conditions ($t\rightarrow \infty$), 
when the currents ${\bf j}_{_N}^\tau (t,{\bf v})$ and ${\bf j}_{_D} 
(t,{\bf v})$ vanish.\\
If $D_{_{N,D}} ({\bf v})=D_{_{N,D}} (v)$ and 
${\bf J}_{_{N,D} }({\bf v})={\bf v} \, {J}_{_{N,D}}(v)/v$ we obtain:  

\begin{equation}
n_{_{N,D}}({\bf v})=\frac{1}{\exp [\beta (E_{_{N,D}}-\mu_{_{N,D}})]-
\kappa_{_{N,D}}} \ \ , 
\end{equation}
\noindent 
where $E_{_{N,D}}=\frac{1}{2} m_{_{N,D}} v^2+V_{_{N,D}} (v)$ and
 
\begin{equation}
\frac{\partial V_{_{N,D}} (v)}{\partial v}=
 \frac{1} {\beta D_{_{N,D}}(v)} \left [J_{_{N,D}}(v)+ 
\frac{\partial D_{_{N,D}} (v)}{\partial v}\right ] - m_{_{N,D}} v \ \ ;
\end{equation}

\noindent 
$\kappa_{_N}=-1$, $\kappa_{_D}=1$, 
$\beta=1/kT$ and $T$ is the temperature of the system, 
$\mu_{_{N,D}}$ is the chemical
potential of the nucleons and the QD particles. 
In Eq. (8), $\partial V_{_{N,D}} (v)/ \partial v$ represents 
the difference between the force acting among the particles 
and the Brownian force $m_{_{N,D}} v$.\\
We note that, in the case of Brownian particles, the potential 
$V_{_{N,D}} (v)$ is a constant 
and $n_{_{N,D}}({\bf v})$ reproduces the standard FD and BE 
statistical distribution.\\
We define the transmutational potential $\eta_{_{2N \rightarrow D}}$ 
as 

\begin{equation}
\frac{r_{_{2N \rightarrow D}}} {r_{_{D \rightarrow 2N}}}=
\exp[\beta\eta_{_{2N \rightarrow D}}]
\end{equation}

\noindent 
and introduce the potential $h(v)$

\begin{equation}
\frac{g_{_N}(v)} {g_{_D}(v)}=\exp[\beta h(v)] \ \ .
\end{equation}

\noindent 
The condition that the net transmutational current, 
given by Eq.(5), must satisfy the equation $j_{_{2N \leftrightarrow D}} 
(\infty,{\bf v})=0$, imposes the two following relations:

\begin{equation}
\eta_{_{2N\rightarrow D}}=\mu_{_D}-2\mu_{_N} \ \ ,
\end{equation}

\begin{equation}
h(v)=V_{_N}^{\tau}(v)+V_{_N}^{\bar\tau}(v)-V_{_D}(v) \ \ .
\end{equation}

\noindent
Equation (11), at any excitation energy $E_{\gamma}$, provides a 
relationship between 
the chemical potentials of the two different types of particles 
and the TP. 
Eq. (12) allows us to obtain the potential $h(v)$ in terms of the 
interaction potentials (constant in the case of brownian 
particles).\\
Let us define the two quantities:

\begin{equation}
\xi_{_{N,D}}=\frac{N_{_{N,D}}}{A} \ \ , 
\end{equation}
\noindent
which are, respectively, the free nucleon fraction and the 
quasi-deuteron fraction, A is the mass number of the nucleus. In other 
words: $\xi_{_N}$ and $2 \xi_{_D}$ represent the probability that 
a nucleon is a free nucleon or a part of a QD. $\xi_{_N}$ and $\xi_{_D}$  
satisfy the normalization condition:

\begin{equation}
\xi_{_N}+2 \xi_{_D}=1 \ \ . 
\end{equation}

\noindent
The fraction $\xi_{_{N,D}}$ of nucleons and of QD is given by:

\begin{equation}
\xi_{_{N,D}} = f_{_{N,D}} \, \frac{1}{\rho} \, 
\left (\frac{m_{_{N,D}}}{h}\right )^3  
\int n_{_{N,D}}({\bf v}) d^3 v \ \ ,
\end{equation}

\noindent
where $\rho$ is the nuclear density and 
$f_{_{N,D}}$ is the spin-isospin degeneration factor 
($f_{_N}=4$ for the nucleons; $f_{_D}=3$ for the QD, having total spin 
$S=1$ and total isospin $T=0$),
$V$ is the nuclear volume, $h$ the Planck constant.\\
Note that we have considered pairs with $L=0$, $S=1$ and $T=0$, 
that are the quantum numbers of a free deuteron. 
Recently, it has been considered the presence of 
QD pairs with $T=1$ to obtain agreement with the experimental results 
on photoabsorption [17-19]. In our case, if we considered 
a QD component with $T=1$ 
and $S=1$, we should have taken the degeneration factor $f_{_D}=6$. 
In that case,  
the value of the QD chemical potential would be reduced, 
because the total number of particles is fixed by the Eq.(15) 
and also the value of the transmutational potential 
$\eta_{_{2N\rightarrow D}}$ would be reduced because of 
Eq.(11). The insertion of this component is not relevant to 
the results obtained in this work.

If, for a given nucleus, $\xi_{_D}$ and $\xi_{_N}$ are known, 
as a function of the excitation energy, inverting Eq.(15), one 
can deduce the chemical 
potentials $\mu_{_N}$ e $\mu_{_D}$ and, by means of equation (11), 
derive the transmutational potential $\eta_{_{2N\rightarrow D}}$ 
as a function of $E_\gamma$.

\section{Calculation of the transmutational potential}

If the energy of the incident photon is fully 
converted into  excitation energy, the nuclear temperature $T$ can be 
calculated from the well known relation:
\begin{equation}
E_{\gamma}=\frac{A}{8} T^2\mbox{  MeV}\ \ .
\end{equation}
\noindent
In our case this relation is not rigorously exact, as  
shown by Montecarlo calculations in an intranuclear cascade model 
[20,21]. Indeed, not all the photon 
energy contributes to increase the nuclear temperature. In the QD 
energy region, the excitation energy is about $15\%$ smaller than the 
photon energy [21].
We have verified that our results are not modified if we consider 
this reduction (for 
a discussion on the validity of Eq.(16) see Ref.[22]).

The QD and nucleon fractions have been determined using our 
QD model of photofission [10]. Let us recall that the following quantities. 
First, 

\begin{equation}
C(N,Z)=7.72 \, \frac{NZ}{A}\ \ ,
\end{equation}

\noindent
is the effective number of QD pairs used to write the photabsorption cross 
section as 
the product of the QD effective number times the photodisintegration 
cross section of the deuteron $\sigma_{_D}$. Second,   

\begin{equation}
F_1(E_{\gamma})=\exp\left (-\frac{D}{E_\gamma}\right )\ \ ,
\end{equation}

\noindent
is the probability that one of the $C(N,Z)$ pairs takes part 
in the photoabsorption reaction (this factor is due to the Pauli 
blocking and the quantity $D/2$ 
is the average energy to excite each nucleon of a 
pair above the 
Fermi level ($D=60 \, MeV$) ). Third, 

\begin{equation}
F_2(E_{\gamma})=\exp\left (-\frac{D-\Gamma(E_\gamma)}
{E_\gamma}\right )\left [ 
1-\exp \left (-\frac{D+\Gamma(E_\gamma)}{E_\gamma}\right)\right ] \ \ ,
\end{equation}

\noindent
is the probability that one of the $C(N,Z)$ pairs 
participates in a photoabsorption evolving toward photofission 
(the product $C(N,Z) F_2 \sigma_{_D}$ is the photofission 
cross section);
$\Gamma(E_\gamma)$ is a polynomial in  
$E_\gamma$ (we have introduced it to reduce, in the first factor of $F_2$, 
the photofission probability at low excitation energies, where 
the photoabsorption without fission is the most probable process, 
and to take into account, by means of the second 
factor, the photofission reduction at high energies). 
In the QD energy region, 
the contribution of the photabsorption to 
the continuum is greater than other contributions. $\Gamma (E_\gamma)$ 
represents the window of energy states, around the Fermi level, allowed 
to be 
occupied by the excited nucleons and pairs to arrange a 
system of particles evolving toward fission.\\ 
From Eq.s (13) and (14) the 
number of QD particles, correctly normalized is:

\begin{eqnarray}
\xi_{_D}(E_\gamma)=\frac{1}{2} F_2(E_\gamma) \ \ , \\  
\xi_{_N}(E_\gamma)=1-F_2(E_\gamma) \ \ .
\end{eqnarray}
\noindent

We have considered three nuclei with very different 
photofission features: $^{209} Bi$, $^{232} Th$, 
$^{238} U$.\\
In the Table the quantity $F_2(E_\gamma)$ for the three different 
nuclei is reported at different excitation energies. 
The numerical values are obtained from the functions 
$\Gamma (E_\gamma)$ given in Ref.[10].

We can calculate the QD and the nucleon fractions, at any excitation 
energy, with Eq.s (20,21). Using Eq.(15), we deduce the QD and nucleon 
chemical potentials and finally, using relation (11), the TP.\\
The nuclear density has been assumed equal to the infinite nuclear 
matter, i.e. $0.17 \, nucleon \, fm^{-3}$.\\
The nuclear interaction of Eq.(12) is described by a square 
well with $V_{_N}=V_{_D}=V_{_0}=-48$ $MeV$. At the low excitation temperature 
$T=1.1 \div 1.9$ $MeV$,
the chemical potentials 
$\mu_{_D}\approx V_{_0}$, for $^{238}U$ and $^{232}Th$ and varies 
between $-55$ $MeV$ and $-51$ $MeV$ for $^{209}Bi$.

The quantity  $\eta_{_{2N\rightarrow D}}(E_\gamma)$ is an increasing 
function of $E_\gamma$, as shown in the Figure. 
As the quantity of $\eta_{_{2N\rightarrow D}}$ 
approaches a positive value, the number of quasi deuterons 
participating to the photofission becomes greater than the 
number of single nucleons. Infact, for $^{238} U$, 
$\eta_{_{2N\rightarrow D}}(90 \, MeV)=0$ and at this energy 
$F_2(90 \, MeV)\approx 0.5$, which equals the value reported in 
the Table. Only in the case of $^{209}Bi$, 
$\eta_{_{2N\rightarrow D}}(E_\gamma)$ is nearly constant 
and negative. Let us note that, when 
$\eta_{_{2N\rightarrow D}}\rightarrow -\infty$, the full system 
is composed by fermions (see Eq. (9)), on the other hand, 
if $\eta_{_{2N\rightarrow D}}\rightarrow +\infty$ the full system 
is composed by bosons.

In ref. [10,11] we have shown that the photofission probability 
$P_f=P_f (E_\gamma)$ is the solution of the following equation 
\begin{equation}
\frac{dP_f}{P_f}=f(\sqrt E_\gamma) \, \frac{d\sqrt E_\gamma}{E_\gamma} \ \ ,
\end{equation}
where $f(\sqrt E_\gamma)$ is a polynomial in $\sqrt E_\gamma$.\\
When $f(\sqrt E_\gamma)$ is nearly constant (as in the case of $^{209} Bi$), 
the solution of Eq.(22) is given by the simple expression 
$P_f=a+b/\sqrt E_\gamma$ which is the result of the statistical model.
Usually $f(\sqrt E_\gamma)$ is not constant and the expression of 
$P_f$ is more involved.\\
Eq.(22) can be seen as a generalized Clausius-Clapeyron 
equation where $f(\sqrt E_\gamma)$ plays, 
in the photofission, the same role of the entalpy in the 
phase transition process. \\
We can assume that the function $f(\sqrt E_\gamma)$ governs 
the first order phase transition from fermions (nucleons) 
to bosons (QD), capable of producing a compound system, 
which then proceeds to a saddle point, to scission and finally to fission.

We can define the nuclear latent heat $dL$ of the transition 
process as
\begin{equation}
dL=\sqrt E_\gamma d\left (\frac{f(\sqrt E_\gamma)}{\sqrt E_\gamma}\right ) \ \ .
\end{equation}
From thermodynamic arguments and taking into account the 
transmutational potential defined in Eq.(9) and (11), the 
latent heat, at the thermodynamic equilibrium, can be written as 
\begin{equation}
L=T\left [  2\, \frac{S_{_N}}{N_{_N}}-\frac{S_{_D}}{N_{_D}}-
\frac{ d\eta_{_{2 N\rightarrow D }} } {dT}  \right ] \ \ ,
\end{equation}
where $S_{_N}$ and $S_{_D}$ are, respectively, the entropies 
of the nucleons and the QD.\\
From Eq.(24), we can realize that the important physical quantity is the 
variation of the transmutational potential with the nuclear temperature 
$T$ (or with $\sqrt E_\gamma)$; infact, this quantity modifies the 
balance of the entropy of the two phases of the system.  
$\eta_{_{2N \rightarrow D}}(T)$ is an increasing 
function (except for $^{209} Bi$, 
where $\eta_{_{2N \rightarrow D}}$ is nearly constant and the 
phase-transition is not relevant), 
therefore $- d\eta_{_{2N \rightarrow D}}/dT$ decreases the latent 
heat, favoring the phase transition or, equivalently, increasing 
the photofission probability.

\section{Microscopic interpretation of the transmutational potential}

Let us now outline a microscopic interpretation of the 
transmutational potential.
In the BCS theory, the ground state is composed by particles all 
paired with opposite spins:

\begin{equation}
\vert BCS>=\prod_{k>0} (u_k+v_k \hat{a}_k ^{\dag}  
\hat{a}_{-k}^{\dag})\vert 0>  \ \ , \end{equation}

\noindent
with the normalization condition:
\begin{equation}
u^2_k+v^2_k=1\ \ .
\end{equation}

\noindent
In Eq.(25), $v^{2}_k$ and $u^{2}_k$ represent, respectively, the probability 
that the state $k$ is occupied by a pair of particles or not. 
We recall that the antisymmetrization is contained in the anticommutation 
properties of the fermionic operators and the product operator 
applies only over the $k>0$ states. The states $k<0$ refers to 
the conjugate states having the third component of the spin with 
opposite values. 

In our case, where QD pairs are contained into the nucleus, 
the state $k$ can be empty or 
occupied by one non-paired nucleon or by a proton-neutron pair 
in a triple spin state. In analogy with Eq.(25), 
we can write the nuclear QD state at a given excitation energy 
in this form:

\begin{equation}
\vert NQD>=\prod_{k>0} (u_k+c_k \hat{a}_k^{\dag}+
	v_k^{^{S=1}} \hat{a}_k^{\dag}  \hat{a}_{-k}^{\dag})\vert 0>  \ \ 
\end{equation}

\noindent
where, now, $-k$ indicates the state with third isospin component 
having the opposite value.\\
The condition in Eq.(26) becomes:

\begin{equation}
u^2_k+c^2_k+v^2_k=1\ \
\end{equation}
\noindent
where $u^{2}_k$ is the probability that the state $k$ is non-occupied; 
$c^{2}_k$ is the probability that the state $k$ is occupied by one 
nucleon; 
$v^{2}_k$ is the probability that the state $k$ is occupied by one 
QD pair.\\
The three coefficients $u_k$, $c_k$, $v_k$ are functions of the 
excitation energy.
With the above definitions we can write:

\begin{equation}
\frac{r_{_{2N\rightarrow D}}}{r_{_{D\rightarrow 2N}}}=\sum_k 
\left (\frac{v_k}{c_k}\right)^2\ \
\end{equation}

\noindent
and using the definition of the transmutational potential 
given in Eq. (9):

\begin{equation}
\eta_{_{2N\rightarrow D}}(E_\gamma)=k_{_B} T \log\left[\sum_k 
\left (\frac{v_k}{c_k}\right )^2\right]\ \ .
\end{equation}

\noindent
This potential contains all the physical information on the 
fermion-boson transmutation valid in the particular context of 
photofission of heavy nuclei, being 
fixed, at the moment, the region between $40$ and $100$ $MeV$. \\
Note that if $v_k=0$ (i.e. if inside the nucleus 
there are no pairs contributing to photofission), the 
transmutational potential from two nucleons to a QD pair 
becomes equal to minus infinite, in agreement with the results 
one can deduce from Eq.(19) of Ref. [10]: all the particles are 
fermions (nucleons).
In conclusion, we remark that the state $\vert NQD>$ of 
Eq.(27) does not represent, as the state $\vert BCS>$ of Eq.(25), 
a redefinition of the vacuum state after having accomplished a 
Bogoliubov transformation; the state $\vert NQD>$  represents an 
excitation state of the nucleus.

\section{Conclusion}

Among the several nucleon pairs that the correlations can arrange 
into nuclei, the quasi-deuterons are important in specific 
processes like, for instance, the photofission in the QD energy 
region. Infact, the empirical-phenomenological QD model describes 
accurately the behavior of the very different photofission cross 
sections of the nuclei $^{209}Bi$, $^{232}Th$ and $^{238}U$.\\
This description can be theoretically justified on the basis of the  
statistical transmutation fermion (nucleon) - boson (QD) due 
to the transmutational potential studied in this work. 
The goal has been accomplished by 
considering, first, a semiclassical kinetic model which describes the  
distribution functions of nucleons (fermions) and quasi-deuterons (bosons) 
and the transmutation from one statistical family to the other one.
These quantities are given in terms of the nucleon and QD chemical 
potentials, of the TP or the transmutation rate and are calculated in 
the framework of the QD model of photofission we introduced in the past and 
applied to the photofission of $^{209}Bi$, $^{232}Th$ and $^{238}U$. 
We have found that the TP $\eta_{_{2N\rightarrow D}}(E_\gamma)$ 
is related to a phase transition in the nuclear system. 
This has be shown by means of a Clausius-Clapeyron equation 
satisfied by the photofission probability.\\
Finally, we have indicated the existence of a relationship between statistical 
models, based on the semiclassical kinetic approach and many-body 
microscopic approaches like the superfluid nuclear model and have shown 
a microscopic interpretation of the transmutational potential.

\vspace{0.5 cm}

\centerline{\bf Acknowledgments}
\noindent
We would like to thank G. Gervino and G. Lapenta for the stimulating 
discussions and for the critical reading of the manuscript.

\vspace{1 cm}


\noindent
{\bf Table Caption}\\
$F_2(E_\gamma)$ for the three nuclei $^{209}Bi$, $^{232}Th$, $^{238}U$.\\


\noindent
{\bf Figure Caption}\\
The transmutational potential as a function of the excitation 
energy for $^{209}Bi$, $^{232}Th$, $^{238}U$ ($V_{_0}=-48$ $MeV$).

\vspace{1 cm}
\centerline{\bf Table}
\vspace{0.5 cm}

$$\offinterlineskip \tabskip=0pt 
\vbox{
\halign 
{\strut  \vrule\vrule#& \quad \hfil # \hfil \quad &
                       \vrule#& \quad \hfil # \hfil &
                       \vrule#& \quad \hfil # \hfil &
                       \vrule#& \quad \hfil # \hfil &
                       \vrule#& \quad \hfil # \hfil &
                       \vrule#& \quad \hfil # \hfil &
                       \vrule#& \quad \hfil # \hfil &
                       \vrule#& \quad \hfil # \hfil &
\vrule\vrule# \cr
\noalign {\hrule}
\noalign {\hrule}
&$E_\gamma  \, [MeV]$ && 40 && 50 && 60 && 70 && 80 && 90 && 100 & \cr
\noalign {\hrule}
\noalign {\hrule}
& $^{209}Bi$ && 0.0002 && 0.0003 && 0.0015 && 0.004 && 0.008 && 0.012 && 
0.019  &\cr
\noalign {\hrule}
& $^{232}Th$ && 0.109 && 0.142 && 0.178 && 0.221 && 0.269 && 0.324 && 0.383  
&\cr
\noalign {\hrule}
& $^{238}U$ && 0.183 && 0.237 && 0.296 && 0.396 && 0.428 && 0.501 && 0.578  
&\cr
\noalign {\hrule}
\noalign {\hrule} }}$$

\end{document}